\documentclass[12pt]{article}

\usepackage[bookmarks=false]{hyperref}  % for \[begin,end]{NoHyper}

\usepackage{mathrsfs}    % for \mathscr{}
\usepackage{mathtools}  % for \prescript{}{}
\usepackage{amsfonts}   % for \mathbb{R}
\usepackage{orcidlink}    % to have a clickable orcid in author.

\usepackage[utf8]{inputenc}
\usepackage[T1]{fontenc}
\usepackage{mathtools}

\newcommand*{\transp}[2][-3mu]{\ensuremath{\mskip1mu\prescript{\smash{\mathrm t\mkern#1}}{}{\mathstrut#2}}}%
\usepackage{amssymb}    % for greater or equal \geqslant.
\usepackage[title]{appendix}
\makeatletter
\newcommand{\bigcdot}{\mathpalette\bigcdot@{.7}}
\newcommand\bigcdot@[2]{\mathbin{\vcenter{\hbox{\scalebox{#2}{$\m@th#1\bullet$}}}}}
\makeatother

\newcommand{\onedot}{\bigcdot}
\newcommand{\twodots}{\bigcdot\bigcdot}
\newcommand{\equpdef}{\mathrel{\stackrel{\makebox[0pt]{\mbox{\normalfont\tiny def}}}{=}}}
\newcommand{\equpdem}{\mathrel{\stackrel{\makebox[0pt]{\mbox{\normalfont\tiny dem}}}{=}}}
\newcommand{\equpcomp}{\mathrel{\stackrel{\makebox[0pt]{\mbox{\normalfont\tiny cas}}}{=}}}

\newcommand{\shorterquad}{\hskip\fontdimen3\font}
\newcommand{\eqdef}{\shorterquad \equpdef \shorterquad}
\newcommand{\eqdem}{\shorterquad \equpdem \shorterquad}
\newcommand{\eqcomp}{\shorterquad \equpcomp \shorterquad}

\newcommand{\myexp}[1]{e^{#1}}
\newcommand{\boxmath}[2][1.1]{\begin{displaymath}\scalebox{#1}{$\displaystyle #2 $}\end{displaymath}}

\newcommand{\boxequa}[2]{\begin{equation}\label{#1}\scalebox{1.1}{$\displaystyle #2 $}\end{equation}}

\begin{document}

\title{\bf Symmetric Lorentz-Minkowski, \\Antisymmetric Dirac-Majorana}
\author{Guy Barrand (\orcidlink{0000-0001-7571-0558})}
%\author{Guy Barrand \shorterquad \orcidlink{0000-0001-7571-0558}}
%\date {\today}
\date {April 6, 2023}
\maketitle

\begin{abstract}
  We show a nice symmetric/antisymmetric relation between the
  four vector Lorentz transformation and the Dirac spinor one in the Majorana
  representation. From the spinor one, we exhibit the antisymmetric pending of the symmetric
  Minkowski metric. We then rewrite the Dirac equation in various ways
  exploiting group properties induced by these relations, and this
  without complex numbers. We show also a nice relation with a five
  dimensional metric. When done, we will see that the traditional
  complex electromagnetic coupling could be handled also without complex numbers
  by just considering two coupled real fields instead of one complex field.
  Finally, we will show that going toward six or ten dimensional spacetime would be more natural
  from a group point of view.
\end{abstract}

\begin{NoHyper}  % to avoid warning related to the production of pdf
                          % bookmarks when section, subsection titles
                          % contain maths.

\section {The coordinate Lorentz transformation, the symmetric
  Minkowski matrix}\label{lorentz_transformation}
A coordinate Lorentz transformation on a tuple $x^{\mu=0,1,2,3}$ of four real numbers can be written:
\boxequa{coordinate_tsf}{
  \mathscr{L}^c(A|x) ^\mu \eqdef (\myexp{A \eta})^\mu_\nu x^\nu
} %end boxequa.
with the Minkowski $\eta$ 4x4 real symmetric matrix being:
\boxmath{
  \eta \eqdef \left(\begin{array}{cccc}1&0&0&0\\0&-1&0&0\\0&0&-1&0\\0&0&0&-1\end{array}\right)
} %end boxmath.
and A being a 4x4 real antisymmetric matrix. Concerning the
exponential, someone must note the
pattern: antisymmetric matrix A, parametrizing a Lorentz
transformation,  multiplied by the constant $\eta$
4x4 real symmetric matrix. A 4x4 real antisymmetric matrix having
six free parameters, we recover the number of parameters (three
rotations plus three boosts) of a Lorentz transformation.

\section {Scalar, vector Lorentz transformations}
 On a field $\phi(x)$ (no index), a tuple $V^{\mu=0,1,2,3}(x)$ of fields, they are defined with:
\boxmath{
  \mathscr{L}^s(A|\phi) ( \mathscr{L}^c(A|x) ) \eqdef \phi(x)
} %end boxmath.
\boxequa{exp_A_eta}{
  \mathscr{L}^{\onedot}(A|V) ^\mu ( \mathscr{L}^c(A|x) ) \eqdef (\myexp{A \eta})^\mu_\nu V^\nu(x)
} %end boxequa.
We can also introduce the ``down'' vector transformation on a $V_{\mu=0,1,2,3}(x)$:
\boxmath{
  \mathscr{L}_{\onedot}(A|V) _\mu ( \mathscr{L}^c(A|x) ) \eqdef
  (\myexp{-A \eta})_\mu^\nu V_\nu(x)
} %end boxmath.

\section {$\eta$ as a metric}
 If the length of a $V^\mu(x)$ is defined, in matrix notation, with:
\boxmath{
  l[\eta](V)(x) \eqdef\transp[0mu]{V} (x) \eta V(x) \eqdef (\transp[0mu]{V}(x))_\mu  \eta^\mu_\nu V^\nu(x)
} %end boxmath.
 ($\transp[0mu]{}$ for the transposition operation), it can be
 shown easily that it is a Lorentz invariant:
\boxmath{
  l[\eta](\mathscr{L}^{\onedot}(A|V) )( \mathscr{L}^c(A|x) ) \eqdem l[\eta](V)(x)
} %end boxmath.
due to the fact that:
\boxmath{
  \prescript{t}{}{ (\myexp{A \eta})} \eta \myexp{A \eta} \eqdem \eta
} %end boxmath.
It is a general property that if A is an antisymmetric
 square matrix and S a symmetric square matrix of same dimension, we have:
\boxmath{
  \prescript{t}{}{(\myexp{A S})} S \myexp{A S} \eqdem \myexp{-S A} \myexp{S A} S \eqdem  S
} %end boxmath.
\boxmath{
  \prescript{t}{}{(\myexp{S A})} A \myexp{S A} \eqdem \myexp{-A S} \myexp{A S} A \eqdem  A
} %end boxmath.
This second property will have some importance in the following. If S and
A are invertible, we have also:
\boxequa{S_inv_invariance}{
  \myexp{A S} S^{-1} \prescript{t}{}{(\myexp{A S})} \eqdem \myexp{A S}
  S^{-1} \myexp{-SA} \eqdem \myexp{A S}
  \myexp{-AS} S^{-1}  \eqdem S^{-1}
} %end boxequa.
\boxmath{
  \myexp{S A} A^{-1} \prescript{t}{}{(\myexp{S A})} \eqdem \myexp{S A}
  A^{-1} \myexp{-AS} \eqdem \myexp{S A}
  \myexp{-SA} A^{-1}  \eqdem A^{-1}
} %end boxmath.

\section {Dirac spinor Lorentz transformation}
On a tuple $\psi^{\alpha=0,1,2,3}(x)$ of fields, we define the ``spinor'' Lorentz
transformation with:
\boxequa{exp_D_A}{
  \mathscr{L}^\circ(A|\psi) ^\alpha ( \mathscr{L}^c(A|x) ) \eqdef (\myexp{\Gamma[A]})^\alpha_\beta \psi^\beta(x)
} %end boxequa.
with $\Gamma[A]$ being:
\boxmath{
  \Gamma[A] \eqdef \frac{1}{8}[\gamma^\mu,\gamma^\nu] (\eta A \eta)^\mu_\nu
} %end boxmath.
with the $\gamma^{\mu=0,1,2,3}$ being four 4x4 complex matrices verifying:
\boxequa{gammas_anticoms}{
  \{ \gamma^\mu,\gamma^\nu \} = 2(\eta^{-1})^\mu_\nu I
} %end boxequa.
with $I$ being the four dimensional identity matrix.

Presented in this form, the $\myexp{\Gamma[A]}$ appearing in formula (\ref{exp_D_A}) is less
appealing than the $\myexp{A\eta}$ appearing in (\ref{exp_A_eta}); it
looks a little bit exotic and much more
complicated, but we are going to show that in fact it can be presented
with the same similar quite simpler structure than in $\myexp{A\eta}$.

About the definition and properties of $\Gamma[A]$, have a look to
Appendix $\ref{appendix_gammas}$ with $S=\eta$. In particular we have the two important properties:
\boxequa{exp_D_gamma_up}{
  \myexp{\Gamma[A]} \gamma^\mu \myexp{-\Gamma[A]} \eqdem (\myexp{-A \eta})^\mu_\nu \gamma^\nu
} %end boxequa.
\boxequa{exp_D_gamma_down}{
  \myexp{\Gamma[A]} \gamma_\mu \myexp{-\Gamma[A]} \eqdem (\myexp{A \eta})_\mu^\nu \gamma_\nu
} %end boxequa.
 with the down $\gamma_\mu$ being:
\boxmath{
   \gamma_\mu \eqdef \eta_\mu^\nu \gamma^\nu
} %end boxmath.

Since the $\myexp{\Gamma[A]}$ matrix looks complex, up so far, the
$\psi^\alpha(x)$ have to be complex numbers too.

\section {The Dirac equation}
 Over a $\psi^\alpha(x)$ tuple, it reads:
\boxmath{
  i (\gamma^\mu)^\alpha_\beta \partial_\mu \psi^\beta(x) = \frac{mc}{\hbar} \psi^\alpha(x)
} %end boxmath.
or in a more compact matrix notation:
\boxequa{Dirac_equation}{
  i \gamma^\mu \partial_\mu \psi(x) = \frac{mc}{\hbar} \psi(x)
} %end boxequa.
By using ($\ref{coordinate_tsf}$), ($\ref{exp_D_A}$) and exploiting ($\ref{exp_D_gamma_up}$) we can
show quite easily that it is a Lorentz invariant.

\section {Terminology, notation}
 In this article, the words "coordinate'', ``scalar'', ``vector'', ``tensor'',
 ``spinor'', etc are used as a qualifier for a transformation and not to
 define a tuple/matrix of numbers or functions/fields as for $x^\mu$,
 $\phi(x)$, $V^{\mu}(x)$, $\psi^{\mu}(x)$ or later $g_{\mu \nu}(x)$.

 On an advanced theory, the same tuple/matrix of numbers/functions may
 be subjected to various transformations, and then labeling them with
 such qualifiers can lead to confusion.  (For example, we can see in Appendix $\ref{appendix_Vierbein}$
 that the tuple of functions $\psi^\alpha(x)$ is subjected to a scalar
 transformation for $\mathscr{R}$ but to a spinor one for the
 $\mathscr{L}_I$ transformation).

 It is only in a context where only one transformation is around that we can displace such
 qualifer toward a tuple/matrix, and then speak for example of the
 $A_\mu(x)$ ``vector'' electromagnetic potential or the $\psi(x)$
 ``spinor''. (In this text, we stick to the tradition of noting a
 tuple subjected to a vector Lorentz transformation $\mathscr{L}^{\onedot}$ with an uppercase latin letter, as $V(x)$,
and of using the lowercase  greek letter $\psi(x)$ for a tuple submitted to
a spinor Lorentz transformation $\mathscr{L}^\circ$).

For the transformations, we use a notation that may look heavy, but
which is in general complete in the sense that it carries all the
needed informations:
\boxmath{
  \mathscr{T}^{qualifier}(parameters|whatever)^{indices}(arguments)
} %end boxmath.
with ``qualifier''' that could be upward or downward, ``parameters'' being a
set of tuples/matrices of numbers/functions and ``whatever'' being a
tuple/matrix of numbers or functions submitted to the transformation. The
result of a transformation being also a tuple/matrix of
numbers/functions, it has in general upward/downward indices and ending brackets with arguments.

For functions/fields depending of parameters, we use:
\boxmath{
  {name}[parameters](arguments)
} %end boxmath.

We use also $\eqdef$ in case an equality is a definition (left side
is defined by the right side), and $\eqdem$ when an equality comes
from a demonstration (left side is demonstrated to be the right
side). We have also $\eqcomp$ in case an equality is demonstrated by
using a computer algebra system (CAS).

We definitely avoid the practice of setting $\hbar=c=1$ which
complicates the reading of the dimensionality of quantities.

Let us go now to the core of this article.

\section {The $A_a$ basis, $\Gamma_a$ matrices and $\theta^a$ parametrization}
 Let us define the antisymmetric $A_{a=1,2,3,4,5,6}$ six real matrices with:
\boxmath{
  A_1 \eqdef \left(\begin{array}{cccc}0&0&0&0\\0&0&0&0\\0&0&0&1\\0&0&-1&0\end{array}\right)
\quad
  A_2 \eqdef \left(\begin{array}{cccc}0&0&0&0\\0&0&0&-1\\0&0&0&0\\0&1&0&0\end{array}\right)
\quad
  A_3 \eqdef \left(\begin{array}{cccc}0&0&0&0\\0&0&1&0\\0&-1&0&0\\0&0&0&0\end{array}\right)
} %end boxmath.
\boxmath{
  A_4 \eqdef \left(\begin{array}{cccc}0&1&0&0\\-1&0&0&0\\0&0&0&0\\0&0&0&0\end{array}\right)
\quad
  A_5 \eqdef \left(\begin{array}{cccc}0&0&1&0\\0&0&0&0\\-1&0&0&0\\0&0&0&0\end{array}\right)
\quad
  A_6 \eqdef \left(\begin{array}{cccc}0&0&0&1\\0&0&0&0\\0&0&0&0\\-1&0&0&0\end{array}\right)
} %end boxmath.
These matrices form a basis for any 4x4 real antisymmetric
matrix (see Appendix $\ref{appendix_gammas}$ for some of their properties).

Instead of parametrizing a Lorentz transformation with a A antisymmetric matrix, we can do it
by using six real numbers $\theta^a$ so that:
\boxmath{
  A = \theta^aA_a \Leftrightarrow\theta^a \eqdem -\frac{1}{2} Tr(A_aA)
} %end boxmath.
With this choice for the $A_a$ matrices, the tuple ($\theta^{1,2,3},0,0,0)$ parametrizes a spatial rotation,
whilst (0,0,0,$\theta^{4,5,6}$) parametrizes a boost.

By using also the $\Gamma_a$ matrices defined with:
\boxmath{
  \Gamma_a \eqdef \frac{1}{8}[\gamma_\mu,\gamma_\nu] (A_a)^\mu_\nu
} %end boxmath.
(see also Appendix $\ref{appendix_gammas}$ with $S=\eta$), we can write now:
\boxmath{
  \mathscr{L}^c(\theta|x) ^\mu \eqdef (\myexp{\theta^a A_a \eta})^\mu_\nu x^\nu
} %end boxmath.
\boxmath{
  \mathscr{L}^{\onedot}(\theta|V) ^\mu (
  \mathscr{L}^c(\theta|x) ) \eqdef (\myexp{\theta^a A_a \eta})^\mu_\nu V^\nu(x)
} %end boxmath.
\boxmath{
  \mathscr{L}^{\circ}(\theta|\psi) ^\alpha (
  \mathscr{L}^c(\theta|x) ) \eqdef (\myexp{\theta^a\Gamma_a})^\alpha_\beta \psi^\beta(x)
} %end boxmath.

\section {Hidden Dirac-Majorana antisymmetric $\xi$ matrix}
 In the Majorana representation, the $\gamma^{\mu=0,1,2,3}$ matrices
 are (see Appendix $\ref{appendix_gammas_majorana}$):
\boxmath{
  \gamma^0 \eqdef i\left(\begin{array}{cccc}0&0&0&-1\\0&0&1&0\\0&-1&0&0\\1&0&0&0\end{array}\right)
\quad
  \gamma^1 \eqdef i\left(\begin{array}{cccc}1&0&0&0\\0&-1&0&0\\0&0&1&0\\0&0&0&-1\end{array}\right)
} %end boxmath.
\boxmath{
  \gamma^2 \eqdef i\left(\begin{array}{cccc}0&0&0&1\\0&0&-1&0\\0&-1&0&0\\1&0&0&0\end{array}\right)
\quad
  \gamma^3 \eqdef i\left(\begin{array}{cccc}0&-1&0&0\\-1&0&0&0\\0&0&0&-1\\0&0&-1&0\end{array}\right)
} %end boxmath.

If now defining the real antisymmetric matrix:
\boxmath{
  \xi \eqdef
  \left(\begin{array}{cccc}0&0&0&1\\0&0&-1&0\\0&1&0&0\\-1&0&0&0\end{array}\right)
  \eqdem i \gamma^0
} %end boxmath.
we can show the remarkable fact that:
\boxmath{
  \Gamma_a \eqdem S_a \xi
} %end boxmath.
with:
\boxmath{
  S_1 \eqdef \frac{1}{2} \left(\begin{array}{cccc}0&-1&0&0\\-1&0&0&0\\0&0&0&1\\0&0&1&0\end{array}\right)
\quad
  S_2 \eqdef \frac{1}{2} \left(\begin{array}{cccc}0&0&-1&0\\0&0&0&-1\\-1&0&0&0\\0&-1&0&0\end{array}\right)
} %end boxmath.
\boxmath{
  S_3 \eqdef \frac{1}{2} \left(\begin{array}{cccc}-1&0&0&0\\0&1&0&0\\0&0&1&0\\0&0&0&-1\end{array}\right)
\quad
  S_4 \eqdef \frac{1}{2} \left(\begin{array}{cccc}1&0&0&0\\0&-1&0&0\\0&0&1&0\\0&0&0&-1\end{array}\right)
} %end boxmath.
\boxmath{
  S_5 \eqdef \frac{1}{2} \left(\begin{array}{cccc}0&0&0&1\\0&0&-1&0\\0&-1&0&0\\1&0&0&0\end{array}\right)
\quad
  S_6 \eqdef \frac{1}{2} \left(\begin{array}{cccc}0&-1&0&0\\-1&0&0&0\\0&0&0&-1\\0&0&-1&0\end{array}\right)
} %end boxmath.
being all real symmetric matrices!

We have the remarkable fact that the coordinate, vector and spinor
Lorentz transformations with the $\theta^a$ as parameters, now look like:
\boxmath{
  \mathscr{L}^c(\theta|x) ^\mu \eqdef (\myexp{\theta^a A_a \eta})^\mu_\nu x^\nu
} %end boxmath.
\boxequa{L_V_eta}{
  \mathscr{L}^{\onedot}(\theta|V) ^\mu (
  \mathscr{L}^c(\theta|x) ) \eqdef (\myexp{\theta^a A_a \eta})^\mu_\nu V^\nu(x)
} %end boxequa.
\boxequa{L_psi_xi}{
  \mathscr{L}^{\circ}(\theta|\psi) ^\alpha (
  \mathscr{L}^c(\theta|x) ) \eqdef (\myexp{\theta^aS_a\xi})^\alpha_\beta \psi^\beta(x)
} %end boxequa.
We note the remarkable and nice interchange of symmetric/antisymmetric
real matrices between
(\ref{L_V_eta}) and (\ref{L_psi_xi}) with:
\boxmath{
  (A_a, \eta) \Leftrightarrow  (S_a, \xi)
\quad \quad
  (antisyms, sym) \Leftrightarrow (syms, antisym)
} %end boxmath.

To spot this interchange, we can write the vector/spinor group commutators together:
\boxequa{group_A_eta}{
  [A_a\eta, A_b\eta] \eqdem {l_{ab}}^c A_c\eta
} %end boxequa.
\boxmath{
  [S_a\xi, S_b\xi] \eqdem {l_{ab}}^c S_c\xi
} %end boxmath.
with the same ${l_{ab}}^c$ real constants (see Appendices
$\ref{appendix_gammas}$ and $\ref{appendix_l_L}$).

We note also that since $\myexp{\theta^aS_a\xi}$ is real, the
$\psi^\alpha(x)$ tuple does not have to be complex and can stay a
priori real.

\section {Four symmetric matrices are lacking}
The set ${S_a}$, despite of being linearly independent, is not a complete set to form
a basis of the set of symmetric 4x4 real matrices $\mathscr{S}(4,
\mathbb{R} )$. We should have ten matrices instead of six;
four are lacking! Four? but we have four $\gamma_\mu$, and if \ldots yes! It can be shown that:
\boxequa{gamma_down_S_xi}{
  \gamma_\mu  \eqdem -i \tilde{S}_\mu \xi
} %end boxequa.
with:
\boxmath{
 \tilde{S}_0 \eqdef I \eqdef \left(\begin{array}{cccc}1&0&0&0\\0&1&0&0\\0&0&1&0\\0&0&0&1\end{array}\right)
\quad
  \tilde{S}_1 \eqdef \left(\begin{array}{cccc}0&0&0&-1\\0&0&-1&0\\0&-1&0&0\\-1&0&0&0\end{array}\right)
} %end boxmath.
\boxmath{
  \tilde{S}_2 \eqdef \left(\begin{array}{cccc}1&0&0&0\\0&1&0&0\\0&0&-1&0\\0&0&0&-1\end{array}\right)
\quad
  \tilde{S}_3 \eqdef \left(\begin{array}{cccc}0&0&-1&0\\0&0&0&1\\-1&0&0&0\\0&1&0&0\end{array}\right)
} %end boxmath.
being all symmetric too!

We note that ($\ref{exp_D_gamma_down}$) can be written:
\boxequa{exp_S_xi_down}{
  \myexp{\theta^a  S_a \xi} \shorterquad (\tilde{S}_\mu \xi)
  \shorterquad  \myexp{-\theta^b  S_b \xi} \eqdem (\myexp{\theta^cA_c \eta})_\mu^\nu \shorterquad (\tilde{S}_\nu \xi)
} %end boxequa.

We note also that we have:
\boxmath{
  \{ \tilde{S}_\mu \xi, \tilde{S}_\nu \xi\} \eqdem -2 \eta^\mu_\nu I
} %end boxmath.

\section{The $\tilde{S}_\alpha\xi$ Dirac equation}
We can rewrite now the Dirac equation as:
\boxequa{S_xi_Dirac_equation}{
  (\eta^{-1})^\mu_\alpha \shorterquad \tilde{S}_\alpha \xi \shorterquad
  \partial_\mu \psi(x) = \frac{mc}{\hbar} \psi(x)
} %end boxequa.
Since all matrices are real, the $\psi^\alpha(x)$ tuple is a priori real.

For reasons that will appear clear later, we are going to rewrite it as:
\boxequa{h_tilde_Dirac_equation}{
  h^{\mu \alpha}(x) \shorterquad \tilde{S}_\alpha \xi \shorterquad \partial_\mu \psi(x) = \frac{mc}{\hbar} \psi(x)
} %end boxequa.
with the $h^{\mu \alpha}(x)$ real fields being constant and defined as:
\boxmath{
  h^{\mu \alpha}(x) \eqdef  h^{\mu \alpha} \eqdef  (\eta^{-1})^\mu_\alpha\eqdem \eta^\mu_\alpha
} %end boxmath.
If the $h^{\mu \alpha}(x)$ fields are Lorentz transformed as:
\boxmath{
  \mathscr{L}^{\twodots}(\theta|h)^{\mu \alpha}
  ( \mathscr{L}^c(\theta|x) ) \eqdef (\myexp{\theta^aA_a\eta})^\mu_\nu
  (\myexp{\theta^bA_b\eta})^\alpha_\beta h^{\nu \beta} (x)
} %end boxmath.
we have, due to ($\ref{S_inv_invariance}$):
\boxmath{
  \mathscr{L}^{\twodots}(\theta|h)^{\mu \alpha} (x) \eqdem (\eta^{-1})^\mu_\alpha
} %end boxmath.
and we can show, with the help of
($\ref{coordinate_tsf}$), ($\ref{L_psi_xi}$), ($\ref{exp_S_xi_down}$),
that ($\ref{h_tilde_Dirac_equation}$) is Lorentz invariant. The introduction of the $h(x)$
fields may look cumbersome, but we do this to be able to compare with
other Dirac like equations coming in further sections.

\section {The $S_a\xi$ more natural Dirac like equation}
Indeed, from the group point of view, since:
\boxmath{
  [S_a\xi, S_b\xi] \eqdem {l_{ab}}^c S_c\xi
} %end boxmath.
and by defining the $L_a$ ``adjoint'' six 6x6 real matrices:
\boxmath{
  (L_a)^b_c \eqdef {l_{ac}}^b
} %end boxmath.
we have the exponential relation:
\boxmath{
  \myexp{\theta^cS_c\xi} \shorterquad S_a\xi \shorterquad \myexp{-\theta^dS_d\xi} \eqdem
  (\myexp{\theta^eL_e})_a^b  \shorterquad S_b\xi
} %end boxmath.
So it would be much more natural to introduce the equation:
\boxequa{h_Dirac_equation}{
  h^{\mu a}(x) \shorterquad S_a \xi \shorterquad \partial_\mu \psi(x) = \frac{mc}{\hbar} \psi(x)
} %end boxequa.
\boxmath{
  \equiv h^{\mu a}(x) \Gamma_a \partial_\mu \psi(x) = \frac{mc}{\hbar} \psi(x)
} %end boxmath.
This equation would be Lorentz invariant if the, a priori real, $h^{\mu a}(x)$ fields be Lorentz transformed as:
\boxequa{h_Lorentz_transform}{
  \mathscr{L}^{\onedot \square}(\theta|h)^{\mu a}
  ( \mathscr{L}^c(\theta|x) ) \eqdef (\myexp{\theta^cA_c\eta})^\mu_\nu
  (\myexp{\theta^d L_d})^a_b h^{\nu b} (x)
} %end boxequa.

It is a remarkable trick from Dirac that the lonely relation
($\ref{gammas_anticoms}$), leading to ($\ref{exp_D_gamma_up}$) avoided
to consider the more natural eqution ($\ref{h_Dirac_equation}$). Once having written
($\ref{exp_D_gamma_up}$), the Dirac equation ($\ref{Dirac_equation}$),
crowded with complex numbers, looks an ad hoc construction from the group point of view. Someone may argue that
($\ref{h_Dirac_equation}$) introduces the extra fields  $h^{\mu a}(x)$
that may look cumbersome, but it is not new to
see the introduction of coworking real fields related to the Dirac equation;
this had been introduced already, through ``vierbein'', by people wanting to handle the Dirac
equation within the framework of general relativity (see Appendix $\ref{appendix_Vierbein}$). We see here that
coworking fields emerged naturally lonely from a group point of view.

\section {The $A_a\eta$ Dirac like equation}
At this point, it is worth noting that we could introduce also the
equation:
\boxmath{
  h^{\mu a}(x) (A_a \eta)^\nu_\lambda \partial_\mu V^\lambda(x) = \frac{mc}{\hbar} V^\nu(x)
} %end boxmath.
\boxequa{h_eta_equation}{
  h^{\mu a}(x) \shorterquad A_a \eta \shorterquad \partial_\mu V(x) = \frac{mc}{\hbar} V(x)
} %end boxequa.
which would be also Lorentz invariant with the same
($\ref{h_Lorentz_transform}$) transformation for the $h(x)$ fields,
the ($\ref{exp_A_eta}$) one for the $V^\mu(x)$ fields, and by exploiting:
\boxmath{
  \myexp{\theta^cA_c\eta} \shorterquad A_a\eta \shorterquad \myexp{-\theta^dA_d\eta} \eqdem
  (\myexp{\theta^eL_e})_a^b  \shorterquad A_b\eta
} %end boxmath.
induced by ($\ref{group_A_eta}$). We are not aware of any attempt to
do physics with such version of the Dirac equation which is invariant
by vector Lorentz transformation only.

\section{The ten $\Sigma_A$ matrices}
We can show that the set of ten matrices $\Sigma_A$ such that:
\boxmath{
  \Sigma_{A=1,10}\eqdef \frac{1}{2} \tilde{S}_{0,1,2,3}, S_{a=1,2,3,4,5,6}
} %end boxmath.
forms a basis of $\mathscr{S}(4, \mathbb{R} )$ with:
\boxmath{
  \{ \Sigma_A \xi \} = \{ \frac{1}{2} i\gamma_\mu , \Gamma_a \}
} %end boxmath.
Concerning group constants we have now:
\boxequa{commutators_Sigmas}{
  [\Sigma_A \xi , \Sigma_B \xi ] \eqdem {C_{AB}}^C \Sigma_C \xi
} %end boxequa.
with the real ${C_{AB}}^C$ that can be computed with the same $Tr()$ technique
as in ($\ref{Tr_l}$). If we define the $C_A$ adjoint ten 10x10 real matrices with:
\boxmath{
  (C_A)^C_B \eqdef {C_{AB}}^C
} %end boxmath.
by using ten $\Theta^A$ parameters, we have the exponential relation:
\boxmath{
  \myexp{\Theta^C \Sigma_C \xi} \shorterquad (\Sigma_A \xi) \shorterquad \myexp{-\Theta^D \Sigma_D
  \xi} \eqdem (\myexp{\Theta^E C_E})_A^B \shorterquad (\Sigma_B \xi)
} %end boxmath.

\subsection {The $\Sigma_A\xi$ Dirac like equation}
If considering that the $\{ \Sigma_A \}$ set is more complete
than the $\{ S_a \}$ set as a basis of 4x4 real symmetric matrices, we can also write:
\boxequa{dirac_Sigmas_xi}{
  h^{\mu A}(x) \shorterquad \Sigma_A \xi \shorterquad \partial_\mu \psi(x) = \frac{mc}{\hbar} \psi(x)
} %end boxequa.
But here, since the associated group involves ten $\Theta^A$
parameters, we can't define a coordinate Lorentz transformation using $\eta$. Instead, we can
continue to speak of an internal transformation $\mathscr{T}$ that let
the $x^\mu$ invariant but transforms $h^{\mu A}(x)$ and $\psi^\alpha(x)$
such that:
\boxmath{
  \mathscr{T}^c(\Theta|x)^\mu \eqdef x^\mu = \delta^\mu_\nu x^\nu
} %end boxmath.
\boxmath{
  \mathscr{T}^{\circ}(\Theta|\psi)^{\alpha} (x) \eqdef (\myexp{\Theta^D \Sigma_D\xi })^\alpha_\beta \psi^\beta (x)
} %end boxmath.
\boxmath{
  \mathscr{T}^{\onedot \square}(\Theta|h)^{\mu A} (x) \eqdef
  (\myexp{\Theta^D C_D} )^A_B h^{\mu B} (x) = \delta^\mu_\nu (\myexp{\Theta^D C_D} )^A_B h^{\nu B} (x)
} %end boxmath.

\section {Connecting to the fifth dimension}
We have a remarkable connection between $\eta$, the ten $\Sigma_A\xi$
4x4 real matrices and a similar logic around a five dimensional 5x5
real matrix metric defined with:
\boxmath{
  \tilde{\eta} \eqdef
  \left(\begin{array}{cc}\eta&0\\0&1\end{array}\right) \eqdef \left(\begin{array}{ccccc}1&0&0&0&0\\0&-1&0&0&0\\0&0&-1&0&0\\0&0&0&-1&0\\0&0&0&0&1\end{array}\right)
} %end boxmath.
and a basis $\tilde{A}_{A=1,10}$ of the ten 5x5 real antisymmetric
matrices defined with:
\boxmath{
  \tilde{A}_1 \eqdef \left(\begin{array}{ccccc}0&0&0&0&1\\0&0&0&0&0\\0&0&0&0&0\\0&0&0&0&0\\-1&0&0&0&0\end{array}\right)
\quad
  \tilde{A}_2 \eqdef \left(\begin{array}{ccccc}0&0&0&0&0\\0&0&0&0&1\\0&0&0&0&0\\0&0&0&0&0\\0&-1&0&0&0\end{array}\right)
} %end boxmath.
\boxmath{
  \tilde{A}_3 \eqdef \left(\begin{array}{ccccc}0&0&0&0&0\\0&0&0&0&0\\0&0&0&0&1\\0&0&0&0&0\\0&0&-1&0&0\end{array}\right)
\quad
  \tilde{A}_4 \eqdef \left(\begin{array}{ccccc}0&0&0&0&0\\0&0&0&0&0\\0&0&0&0&0\\0&0&0&0&1\\0&0&0&-1&0\end{array}\right)
} %end boxmath.
\boxmath{
  \tilde{A}_{4+a(=1,2,3,4,5,6)=5,6,7,8,9,10} \eqdef \left(\begin{array}{cc}A_a&0\\0&0\end{array}\right)
} %end boxmath.
It appears that:
\boxmath{
  [ \tilde{A}_A \tilde{\eta} , \tilde{A}_B \tilde{\eta} ] \eqdem {C_{AB}}^C \tilde{A}_C \tilde{\eta}
} %end boxmath.
with the same group constants as in
($\ref{commutators_Sigmas}$): nicely strange!
Since the number of parameters in a (n=4)x4 symmetric matrix is
$\frac{n(n+1)}{2}=10$ and the number of parameters for a (d=n+1=5)x5
antisymmetric one is $\frac{d(d-1)}{2}=\frac{(n+1)n}{2}=10$, it is not surprising
to find the same number of parameters, but it is more surprising to
see that both set of ten 4x4 matrices $\{ \Sigma_A\xi \}$ and the upper ten 5x5
$\{ \tilde{A}_A\tilde{\eta} \}$ matrices give exactly
the same group constants.

\subsection {The five dimensional Dirac like equation}
If, in the below, the index $\mu$ and $\nu$=0,1,2,3 are taken for A=1,2,3,4 and
a,b,c=1,2,3,4,5,6 for A=5,6,7,8,9,10, we can check that:
\boxmath{
  [ \tilde{A}_a \tilde{\eta} \tilde{A}_b \tilde{\eta} ] \eqdem {l_{ab}}^c \tilde{A}_c \tilde{\eta}
} %end boxmath.
\boxmath{
  [\tilde{A}_a \tilde{\eta}, \tilde{A}_\mu \tilde{\eta}] \eqdem ( A_a \eta)_\mu^\nu \tilde{A}_\nu \tilde{\eta}
} %end boxmath.
We have then, with 5x5 real matrices:
\boxmath{
  \myexp{ \theta^a \tilde{A}_a \tilde{\eta}}  \shorterquad
  (\tilde{A}_\mu \tilde{\eta}) \shorterquad
  \myexp{-\theta^b \tilde{A}_b \tilde{\eta}} \eqdem (\myexp{
    \theta^cA_c \eta })_\mu^\nu \shorterquad (\tilde{A}_\nu \tilde{\eta})
} %end boxmath.
 similar to ($\ref{exp_S_xi_down}$), so that we can write an equation
 similar to the Dirac one but on a $\Psi^{\alpha=0,1,2,3,5} (x)$ tuple:
\boxequa{Dirac_5D}{
  (\eta^{-1})^\mu_\alpha \shorterquad \tilde{A}_\alpha \tilde{\eta}
  \shorterquad \partial_\mu \Psi(x) = \frac{mc}{\hbar} \Psi(x)
} %end boxequa.
with $\Psi(x)$ transforming in a Lorentz transformation with:
\boxmath{
  \mathscr{L}^{5\onedot}(\theta|\Psi) ^\alpha (
  \mathscr{L}^c(\theta|x) ) \eqdef (\myexp{\theta^a \tilde{A}_a \tilde{\eta}})^\alpha_\beta \Psi^\beta(x)
} %end boxmath.
In this equation there is no exotic complex gamma matrices and the
spinor transformation in four dimensions becomes a more natural vector
transformation in five dimensions!

\section {A word on electromagnetism}
 Thanks to the Majorana representation of the gamma matrices, we have
 been able to get rid of complex numbers in our various
 equations, but traditionally the sticky ``$i$'' appears also when handling
 electromagnetism by writing:
\boxequa{Dirac_equation_EM}{
  i \gamma^\mu \{ \partial_\mu \psi(x) +i \frac{q}{\hbar c} \Phi_\mu(x) \psi(x)\} = \frac{mc}{\hbar} \psi(x)
} %end boxequa.
with q being the electromagnetic charge and $\Phi_\mu(x)$ being the
electromagnetic potential related to the three dimensional
Maxwell real $U(t,\vec{x})$ and $\vec{A}(t,\vec{x})$ with:
\boxmath{
   \Phi_\mu(x\eqdef(ct,\vec{x})) \eqdef (U(t,\vec{x}),-\vec{A}(t,\vec{x}))
} %end boxmath.
The $\Phi_\mu(x)$ would be Lorentz transformed with:
\boxmath{
  \mathscr{L}_{\onedot}(A|\Phi) _\mu ( \mathscr{L}^c(A|x) ) \eqdef (\myexp{-A \eta})_\mu^\nu \Phi_\nu(x)
} %end boxmath.

 By using our $\tilde{S}_\alpha\xi$  representation, it becomes:
\boxequa{S_xi_Dirac_equation_EM}{
  h^{\mu \alpha} \shorterquad \tilde{S}_\alpha \xi \shorterquad \{ \partial_\mu \psi(x) +i
  \frac{q}{\hbar c} \Phi_\mu(x) \psi(x) \} = \frac{mc}{\hbar} \psi(x)
} %end boxequa.
which exhibits the fact that ``$i$'' appears now only in the
electromagnetic coupling to $\Phi_\mu(x)$. The complex coupling induces
that $\psi(x)$ has to be complex, but if writing:
\boxmath{
  \psi[\mathscr{V},\mathscr{W}](x) \eqdef \mathscr{V}(x) + i \mathscr{W}(x)
} %end boxmath.
we see that ($\ref{S_xi_Dirac_equation_EM}$) can be written as two
equations without complex numbers on two real coupled fields $\mathscr{V}(x)$ and $\mathscr{W}(x)$:

\boxmath{
  h^{\mu \alpha} \shorterquad \tilde{S}_\alpha \xi \shorterquad \{ \partial_\mu \mathscr{V}(x) -
\frac{q}{\hbar c} \Phi_\mu(x) \mathscr{W}(x) \} = \frac{mc}{\hbar} \mathscr{V}(x)
} %end boxmath.
\boxmath{
  h^{\mu \alpha} \shorterquad \tilde{S}_\alpha \xi \shorterquad \{ \partial_\mu \mathscr{W}(x) +
  \frac{q}{\hbar c} \Phi_\mu(x) \mathscr{V}(x) \} = \frac{mc}{\hbar} \mathscr{W}(x)
} %end boxmath.
or:
\boxequa{Dirac_EM_V}{
  [ h^{\mu \alpha} \shorterquad \tilde{S}_\alpha \xi \shorterquad \partial_\mu -
\frac{mc}{\hbar} I ] \mathscr{V}(x) = [\frac{q}{\hbar c}  \shorterquad
\Phi_\mu(x) \shorterquad
h^{\mu \alpha} \shorterquad \tilde{S}_\alpha \xi
] \mathscr{W}(x)
} %end boxequa.
\boxequa{Dirac_EM_W}{
  [ h^{\mu \alpha} \shorterquad \tilde{S}_\alpha \xi \shorterquad \partial_\mu -
\frac{mc}{\hbar} I ] \mathscr{W}(x) = - [\frac{q}{\hbar c} \shorterquad
 \Phi_\mu(x) \shorterquad h^{\mu \alpha} \shorterquad \tilde{S}_\alpha \xi]
 \mathscr{V}(x)
} %end boxequa.

We see now that complex numbers appear for electromagnetism in ($\ref{Dirac_equation_EM}$)
mainly to write in a more compact form two coupled real quantities.

\subsection {Charge conjugation transformation $\mathscr{C}$}
With ($\ref{Dirac_equation_EM}$), charge conjugation transformation would consist to find a transformation
$\mathscr{C}(\psi)$
such that:
\boxequa{Dirac_equation_EM_CC}{
  i \gamma^\mu \{ \partial_\mu - i \frac{q}{\hbar c} \Phi_\mu(x) \}
  \mathscr{C}(\psi) (x)= \frac{mc}{\hbar} \mathscr{C}(\psi) (x)
} %end boxequa.
By using the Dirac (or Chiral) representation of the gamma matrices, we get that:
\boxequa{CC_Dirac}{
  \mathscr{C}(\psi) (x) \eqdem i \gamma^2 \psi^*(x)
} %end boxequa.
but by working with ($\ref{Dirac_EM_V}, \ref{Dirac_EM_W}$), we see that:
\boxequa{CC_Dirac_V_W}{
  \mathscr{C}(\mathscr{V},\mathscr{W}) (x) \eqdem (\mathscr{W},\mathscr{V}) (x)
} %end boxequa.
that is to say the charge conjugation is reduced to just a swapping of
the two coupled fields! We find this much more appealing than the much
more algebraic ($\ref{CC_Dirac}$). (On $\psi(x)$, ($\ref{CC_Dirac_V_W}$) would be written
$\mathscr{C}(\psi)(x) \eqdem i \psi^*(x)$).

\subsection {$\mathscr{C}$ transformation and our other Dirac like equations}
The same conclusion would be reached with our other equations
($\ref{h_Dirac_equation}$), ($\ref{h_eta_equation}$),
($\ref{dirac_Sigmas_xi}$) and even with the five dimensioal
($\ref{Dirac_5D}$) one. Here too, electromagnetism could be handled by
two coupled real fields.

\subsection {About Maxwell equations}
When writing Maxwell equations in a four dimensional form, it appears
equations like:
\boxmath{
  \partial_\mu F^{\mu \nu} (x)= J^\nu(x)
} %end boxmath.
where $F^{\mu \nu} (x)$ is antisymmetric. This equation is Lorentz
invariant with:
\boxmath{
  \mathscr{L}^{\onedot}(\theta|J)^\mu ( \mathscr{L}^c(\theta|x) ) \eqdef (\myexp{\theta^aA_a \eta})^\mu_\nu J^\nu(x)
} %end boxmath.
\boxequa{tsf_F}{
  \mathscr{L}^{\twodots}(\theta|F)^{\mu \nu}( \mathscr{L}^c(\theta|x) ) \eqdef
  (\myexp{\theta^aA_a \eta})^\mu_\alpha (\myexp{\theta^bA_b \eta})^\nu_\beta F^{\alpha \beta}(x)
} %end boxequa.

 Since $F^{\mu \nu}$ is antisymmetric, we can introduce the six
 $F^a(x)$ fields such that:
\boxmath{
  F^{\mu \nu}(x) \eqdef (A_a)^\mu_\nu F^a(x)
} %end boxmath.

With this, the upper equation becomes:
\boxequa{A_a_F}{
  (A_a)^\mu_\nu \partial_\mu F^a (x) = J^\nu(x)
} %end boxequa.
which reminds the ($\ref{h_eta_equation}$) equation.

Since (seee Appendix $\ref{appendix_gammas}$ with $S=\eta$) we have:
\boxmath{
  \myexp{\theta^cA_c \eta} A_a \prescript{t}{} (\myexp{\theta^dA_d \eta}) \eqdem
  (\myexp{\theta^eL_e})_a^b A_b
} %end boxmath.

We can show that ($\ref{tsf_F}$) induces:
\boxmath{
  \mathscr{L}^{\square}(\theta|F)^a( \mathscr{L}^c(\theta|x) ) \eqdem
  (\myexp{\theta^c L_c})^a_b F^b(x)
} %end boxmath.
which let ($\ref{A_a_F}$) Lorentz invariant.

\section {Toward six or ten dimensional "adjoint'' spacetime?}
In the case of ($\ref{S_xi_Dirac_equation}$) and ($\ref{Dirac_5D}$), $h(x)$ is a constant and a
square (4x4) real matrix, but in case of ($\ref{h_Dirac_equation}$) , ($\ref{h_eta_equation}$) and
($\ref{dirac_Sigmas_xi}$) it is not a square matrix (4x6 or 4x10) and a priori not a
constant field. Being not square is not handy, but we can recover that
by just going to a space-time with six or ten dimensions! Indeed, if
instead of ($x^{\mu=0,1,2,3}$) coordinates we go toward  ($X^{a=1 \rightarrow 6}$)
or ($X^{A=1 \rightarrow 10}$), then ($\ref{h_Dirac_equation}$) or
($\ref{dirac_Sigmas_xi}$) become:
\boxmath{
  h^{a b}(X) \shorterquad S_b \xi \shorterquad \partial_a \psi(X) = \frac{mc}{\hbar} \psi(X)
} %end boxmath.
\boxmath{
  h^{A B}(X) \shorterquad \Sigma_B \xi \shorterquad \partial_A \psi(X) = \frac{mc}{\hbar} \psi(X)
} %end boxmath.
They will be invariant with the transformations:
\boxmath{
  \mathscr{T}^c(\theta|X)^a  \eqdef (\myexp{\theta^d L_d})^a_b X^b
} %end boxmath.
\boxmath{
  \mathscr{T}^\circ(\theta|\psi)^\alpha
  ( \mathscr{T}^c(\theta|X )) \eqdef (\myexp{\theta^a S_a \xi})^\alpha_\beta
  \psi^\beta (X)
} %end boxmath.
\boxmath{
  \mathscr{T}^{\square \square}(\theta|h)^{a b}
  ( \mathscr{T}^c(\theta|X )) \eqdef (\myexp{\theta^cL_c})^a_e
  (\myexp{\theta^d L_d})^b_f h^{e f} (X)
} %end boxmath.
or:
\boxmath{
  \mathscr{T}^c(\Theta|X)^A  \eqdef (\myexp{\Theta^D C_D})^A_B X^B
} %end boxmath.
\boxmath{
  \mathscr{T}^\circ(\Theta|\psi)^\alpha
  ( \mathscr{T}^c(\Theta|X )) \eqdef (\myexp{\Theta^A \Sigma_A \xi})^\alpha_\beta
  \psi^\beta (X)
} %end boxmath.
\boxmath{
  \mathscr{T}^{\square \square}(\Theta|h)^{A B}
  ( \mathscr{T}^c(\Theta|X )) \eqdef (\myexp{\Theta^CC_C})^A_E
  (\myexp{\Theta^D C_D})^B_F h^{E F} (X)
} %end boxmath.

We can even show that we can have constant h(X) fields. Indeed, if taking:
\boxmath{
  h^{a b}(X) \eqdef G^a_b
} %end boxmath.
with the $G$ 6x6 real matrix defined by:
\boxmath{
  G \eqdef G_6 \eqdef
  \left(\begin{array}{cc}I_{3x3}&0\\0&-I_{3x3}\end{array}\right)
\quad
  I_{3x3} \eqdef \left(\begin{array}{ccc}1&0&0\\0&1&0\\0&0&1\end{array}\right)
} %end boxmath.
we have:
\boxmath{
  \mathscr{T}^{\square \square}(\theta|h)^{a b}
  ( \mathscr{T}^c(\theta|X )) \eqdem h^{a b}
} %end boxmath.
and the same for $h^{A B}(X)$ with the $G_{10}$ 10x10 real matrix defined by:
\boxmath{
  h^{A B}(X) \eqdef (G_{10})^A_B
\quad
  G_{10} \eqdef \left(\begin{array}{cc} \eta &0\\0&G_6 \end{array}\right)
} %end boxmath.

\section {Checked numerically by computer}
 All the formulas used in this article had been checked numerically
 by computer. On the author \href{https://github.com/gbarrand}{GitHub gbarrand}, the
 repository ``papers'' contains the open source ``SMAD'' C++ unitary
 test program that verifies the formulas found here.

\section {Conclusions}
Thanks to E.Majorana, it is interesting to see that the four
dimensional vector and spinor Lorentz transformations show a similar ``AS/SA'' real matrix pattern  (S=Symmetric, A=Antisymmetric), and that the
``SA'' real pattern for the spinor transformation is connected to an ``AS'' real pattern in fifth
dimension. With this in head, we see now the Dirac spinor Lorentz
transformation differently; instead of being related to some exotic mathematical trick through
complex Dirac matrices introduced through ($\ref{gammas_anticoms}$),
we see it now deeply related to a real antisymmetric $\xi$ matrix
which is the pending of the real symmetric $\eta$ matrix for the four
vector transformation. By pushing to the fifth
dimension, we can even see it as a five vector transformation.
Moreover, following the logic around this $\xi$ matrix, we see that
the traditional Dirac equation looks rather incomplete and could be
extended in a more natural way to something as
($\ref{h_Dirac_equation}$) or ($\ref{dirac_Sigmas_xi}$)  that we write again here:
\boxmath{
  h^{\mu a}(x) \shorterquad S_a \xi \shorterquad \partial_\mu \psi(x) = \frac{mc}{\hbar} \psi(x)
} %end boxmath.
\boxmath{
  h^{\mu A}(x) \shorterquad \Sigma_A \xi \shorterquad \partial_\mu \psi(x) = \frac{mc}{\hbar} \psi(x)
} %end boxmath.
with the $\psi(x)$ tuple not needed to be complex at this point, and the
coworking $h(x)$ real fields coming naturally from a group point of
view. We saw also that electromagnetism could be handled without complex
numbers in all our Dirac like equations by just introducing two
coupled real fields. Finally, we showed that
going toward six or ten dimensional spacetime would be more natural
from a group point of view.

%/////////////////////////////////////////////////////////////
%/// Appendix: ////////////////////////////////////////////////
%/////////////////////////////////////////////////////////////
\begin{appendices}

\section {$\gamma^\mu$  in Majorana representation}\label{appendix_gammas_majorana}
 The $\gamma^{\mu=0,1,2,3}$ in the Majorana representation are presented
 in general as (see \cite{ref_iz} p.694):
\boxmath{
  \gamma^0 \eqdef \left(\begin{array}{cc}0&\sigma_2\\\sigma_2&0\end{array}\right)
\quad
  \gamma^1 \eqdef \left(\begin{array}{cc}i\sigma_3&0\\0&i\sigma_3\end{array}\right)
} %end boxmath.
\boxmath{
  \gamma^2 \eqdef \left(\begin{array}{cc}0&-\sigma_2\\\sigma_2&0\end{array}\right)
\quad
  \gamma^3 \eqdef \left(\begin{array}{cc}-i\sigma_1&0\\0&-i\sigma_1\end{array}\right)
} %end boxmath.
with the standard Pauli 2x2 complex matrices being defined as:
\boxmath{
  \sigma_1 \eqdef \left(\begin{array}{cc}0&1\\1&0\end{array}\right)
\quad \quad
  \sigma_2 \eqdef \left(\begin{array}{cc}0&-i\\i&0\end{array}\right)
\quad \quad
  \sigma_3 \eqdef \left(\begin{array}{cc}1&0\\0&-1\end{array}\right)
} %end boxmath.
With less complex numbers around, we can write them directly as:
\boxmath{
  \gamma^0 \eqdef i\left(\begin{array}{cccc}0&0&0&-1\\0&0&1&0\\0&-1&0&0\\1&0&0&0\end{array}\right)
\quad
  \gamma^1 \eqdef i\left(\begin{array}{cccc}1&0&0&0\\0&-1&0&0\\0&0&1&0\\0&0&0&-1\end{array}\right)
} %end boxmath.
\boxmath{
  \gamma^2 \eqdef i\left(\begin{array}{cccc}0&0&0&1\\0&0&-1&0\\0&-1&0&0\\1&0&0&0\end{array}\right)
\quad
  \gamma^3 \eqdef i\left(\begin{array}{cccc}0&-1&0&0\\-1&0&0&0\\0&0&0&-1\\0&0&-1&0\end{array}\right)
} %end boxmath.

\section {$\gamma_\mu$, $\Gamma[A]$ matrices}\label{appendix_gammas}
Let us have $\gamma_\mu$ and a S real symmetric square matrices such that:
\boxmath{
  \{ \gamma_\mu,\gamma_\nu \} \eqdef 2S^\mu_\nu I
} %end boxmath.
$I$ being the identity matrix. Note that we do not specify the dimensions of them! The $\gamma_\mu$
can be real or complex.
With a little bit of algebra, we can show that it induces:
\boxequa{anticom_anticom_gammas}{
  [ [\gamma_\mu,\gamma_\nu] , \gamma_\alpha ] \eqdem 4 S^\alpha_\nu
  \gamma_\mu - 4 S^\alpha_\mu \gamma_\nu
} %end boxequa.
If A being a real antisymmetric square matrix of same dimension as S, let us define $\Gamma[A]$:
\boxmath{
  \Gamma[A] \eqdef \frac{1}{8}[\gamma_\mu,\gamma_\nu] (A)^\mu_\nu
} %end boxmath.
The relation ($\ref{anticom_anticom_gammas}$) induces:
\boxmath{
   [\Gamma[A] , \gamma_\mu ] \eqdem (A S)_\mu^\nu \gamma_\nu
 } %end boxmath.
It is a general property (see \cite{ref_iz} p.70) that for square matrices M, $N_a$ of same dimension:
\boxequa{M_N_a_down}{
  [ M, N_a ] \eqdef C_a^b N_b \quad \Rightarrow \quad \myexp{M} N_a \myexp{-M} \eqdem (\myexp{C})_a^b N_b
} %end boxequa.
Then we have the important property:
\boxmath{
  \myexp{\Gamma[A]} \gamma_\mu \myexp{-\Gamma[A]} \eqdem (\myexp{A S})_\mu^\nu \gamma_\nu 
} %end boxmath.

\subsection {$\gamma^\mu$  ``up'' matrices}
 If defining $\gamma^\mu$ with:
\boxmath{
   \gamma^\mu \eqdef = (S^{-1})^\mu_\nu \gamma_\nu
} %end boxmath.
we have:
\boxmath{
  \{ \gamma^\mu,\gamma^\nu \} \eqdem 2(S^{-1})^\mu_\nu I
} %end boxmath.
\boxmath{
  \Gamma[A] \eqdem \frac{1}{8}[\gamma^\mu,\gamma^\nu] (SAS)^\mu_\nu
} %end boxmath.
\boxmath{
   [\Gamma[A] , \gamma^\mu ] \eqdem (-A S)^\mu_\nu \gamma^\nu
 } %end boxmath.
It is a general property (see \cite{ref_iz} p.70) that for square matrices
M, $N^a$ of same dimension:
\boxmath{
  [ M, N^a ] \eqdef C^a_b N^b \quad \Rightarrow \quad \myexp{M} N^a \myexp{-M} \eqdem (\myexp{C})^a_b N^b
} %end boxmath.
Then we have the important property:
\boxmath{
  \myexp{\Gamma[A]} \gamma^\mu \myexp{-\Gamma[A]} \eqdem (\myexp{-A S})^\mu_\nu \gamma^\nu
} %end boxmath.

\subsection {The $A_a$ basis}
 Let us define the antisymmetric $A_{a=1,2,3,4,5,6}$ six matrices with:
\boxmath{
  A_1 \eqdef \left(\begin{array}{cccc}0&0&0&0\\0&0&0&0\\0&0&0&1\\0&0&-1&0\end{array}\right)
\quad
  A_2 \eqdef \left(\begin{array}{cccc}0&0&0&0\\0&0&0&-1\\0&0&0&0\\0&1&0&0\end{array}\right)
\quad
  A_3 \eqdef \left(\begin{array}{cccc}0&0&0&0\\0&0&1&0\\0&-1&0&0\\0&0&0&0\end{array}\right)
} %end boxmath.
\boxmath{
  A_4 \eqdef \left(\begin{array}{cccc}0&1&0&0\\-1&0&0&0\\0&0&0&0\\0&0&0&0\end{array}\right)
\quad
  A_5 \eqdef \left(\begin{array}{cccc}0&0&1&0\\0&0&0&0\\-1&0&0&0\\0&0&0&0\end{array}\right)
\quad
  A_6 \eqdef \left(\begin{array}{cccc}0&0&0&1\\0&0&0&0\\0&0&0&0\\-1&0&0&0\end{array}\right)
} %end boxmath.
These matrices form a basis for any 4x4 real antisymmetric matrix. We can show that:
\boxmath{
  Tr(A_aA_b) \eqdem -2 \delta_{ab}
} %end boxmath.

\subsection {$A_aS$, adjoint matrices}
It appears that:
\boxmath{
  A_aSA_b-A_bSA_a
} %end boxmath.
is antisymmetric, and then we can develop it on the $A_a$ basis and write:
\boxequa{AS_commutator}{
   A_aSA_b-A_bSA_a \eqdem {C_{ab}}^c A_c \Rightarrow \quad [A_aS, A_b
   S] \eqdem {C_{ab}}^c A_c S
} %end boxequa.

The ${C_{ab}}^c$ can be computed by using the matrix trace $Tr()$ with:
\boxequa{Tr_l}{
  Tr[ A_d(A_a S A_b- A_b S A_a) ] = {C_{ab}}^c Tr(A_d A_c) = {C_{ab}}^c (-2\delta_{dc}) \eqdem -2 {C_{ab}}^d
} %end boxequa.
By defining the $C_a$ ``adjoint'' matrices:
\boxmath{
  (C_a)^b_c \eqdef {C_{ac}}^b
} %end boxmath.
we have the commutators:
\boxmath{
  [C_a,C_b] \eqdem {C_{ab}}^c C_c
} %end boxmath.
and by exploiting ($\ref{M_N_a_down}$) we have:
\boxmath{
  \myexp{\theta^cA_cS} \shorterquad A_a S \shorterquad
  \myexp{-\theta^dA_dS} \eqdem (\myexp{\theta^eC_e})_a^b \shorterquad
  A_b S
} %end boxmath.
\boxmath{
  \myexp{\theta^cC_c} \shorterquad C_a \shorterquad \myexp{-\theta^dC_d} \eqdem (\myexp{\theta^eC_e})_a^b \shorterquad C_b
} %end boxmath.
If S is invertible, from the first upper relation we can deduce that we have also:
\boxmath{
  \myexp{\theta^cA_cS} A_a \prescript{t}{} (\myexp{\theta^dA_dS}) \eqdem
  (\myexp{\theta^eC_e})_a^b A_b
} %end boxmath.

\subsection {$\Gamma_a$ matrices}
By using the upper $A_a$ real matrices, we define:
\boxmath{
  \Gamma_a \eqdef \frac{1}{8}[\gamma_\mu,\gamma_\nu] (A_a)^\mu_\nu
} %end boxmath.
We can demonstrate that the commutators ($\ref{AS_commutator}$) induce:
\boxmath{
  [\Gamma_a,\Gamma_b] \eqdem {C_{ab}}^c \Gamma_c
} %end boxmath.
with the same real group constants ${C_{ab}}^c$. By exploiting
($\ref{M_N_a_down}$) we have also:
\boxmath{
  \myexp{\theta^c\Gamma_c} \Gamma_a \myexp{-\theta^d\Gamma_d} \eqdem (\myexp{\theta^eC_e})_a^b \Gamma_b
} %end boxmath.

\subsection {The Baker–Campbell–Hausdorff development}
If X and Y are matrices of same dimension, we have:
\boxmath{
  \myexp{X} \myexp{Y} = \myexp{Z}
} %end boxmath.
with:
\boxmath{
  Z[X,Y] \eqdem X+Y+\frac{1}{2}[X,Y]+\frac{1}{12} [X,[X,Y]] - \frac{1}{12} [Y,[X,Y]]+...
} %end boxmath.

\subsection {The $\oplus$ group operation}
We can then define the $\oplus$ group operation over some
$\theta_1^a$ and $\theta_1^a$ parameters such
that:
\boxmath{
  (\theta_1 \oplus \theta_2)^a \eqdem \theta_1^a+\theta_2^a+\frac{1}{2}
  {C_{bc}}^a \theta_1^b \theta_2^c
+\frac{1}{12} {C_{ed}}^a {C_{bc}}^d \theta_1^e \theta_1^b \theta_2^c
-\frac{1}{12} {C_{ed}}^a {C_{bc}}^d \theta_2^e \theta_1^b \theta_2^c
+...
} %end boxmath.
following a Baker–Campbell–Hausdorff development. It permits to have:
\boxmath{
  \myexp{\theta_1^a A_aS}  \myexp{\theta_2^b A_bS} \eqdem \myexp{ (\theta_1\oplus \theta_2)^a A_aS}
} %end boxmath.
\boxmath{
  \myexp{\theta_1^a \Gamma_a}  \myexp{\theta_2^b \Gamma_b} \eqdem \myexp{(\theta_1\oplus \theta_2)^a \Gamma_a}
} %end boxmath.
which enforce that $\myexp{AS}$ and $\myexp{\Gamma[A]}$ are two representations of the same group.

\section{ ${l_{ab}}^c$ and    $L_a$ matrices}\label{appendix_l_L}
With the real ${l_{ab}}^c$ introduced through:
\boxmath{
  [A_a\eta, A_b\eta] \eqdem {l_{ab}}^c A_c\eta
} %end boxmath.
and the $L_a$ adjoint 6x6 real matrices defined with:
\boxmath{
  (L_a)^b_c \eqdef {l_{ac}}^b
} %end boxmath.
we have:
\boxmath{
  L_{j=1,2,3} \eqdem \left(\begin{array}{cc}-\epsilon_j&0\\0&-\epsilon_j\end{array}\right)
\quad
  L_{a=4,5,6=j+3} \eqdem \left(\begin{array}{cc}0&\epsilon_j\\-\epsilon_j&0\end{array}\right)
} %end boxmath.
with:
\boxmath{
  \epsilon_1 \eqdef \left(\begin{array}{ccc}0&0&0\\0&0&1\\0&-1&0\end{array}\right)
\quad \quad
  \epsilon_2 \eqdef \left(\begin{array}{ccc}0&0&-1\\0&0&0\\1&0&0\end{array}\right)
\quad \quad
  \epsilon_3 \eqdef \left(\begin{array}{ccc}0&1&0\\-1&0&0\\0&0&0\end{array}\right)
} %end boxmath.
 The $L_a$ have all the properties shown in the Appendix \ref{appendix_gammas} with
 $S=\eta$, In particular we have:
\boxmath{
  \myexp{\theta^cA_c\eta} \shorterquad A_a\eta \shorterquad
  \myexp{-\theta^dA_d\eta} \eqdem (\myexp{\theta^eL_e})_a^b
  \shorterquad  A_b\eta
} %end boxmath.
\boxmath{
  \myexp{\theta^c\Gamma_c} \shorterquad \Gamma_a \shorterquad
  \myexp{-\theta^d\Gamma_d} \eqdem (\myexp{\theta^eL_e})_a^b
  \shorterquad \Gamma_b
} %end boxmath.
\boxmath{
  \myexp{\theta^cL_c} \shorterquad L_a \shorterquad
  \myexp{-\theta^dL_d} \eqdem (\myexp{\theta^eL_e})_a^b \shorterquad L_b
} %end boxmath.

\section {$\mathscr{R}$  transformation, Vierbein}\label{appendix_Vierbein}
\subsection {General frame transformation $\mathscr{R}$}
One feature of general relativity is the invariance of
formulas according to a general frame transformation. The idea behind
this being that laws of physics should look the same in any reference frame. 
With the same kind of notation used in the first paragraph
($\ref{lorentz_transformation}$), a general coordinate transformation
on the tuple $x^{\mu=0,1,2,3}$ can be written:
\boxmath{
  \mathscr{R}^c(r|x) ^\mu \eqdef r^\mu(x)
} %end boxmath.
the $r^\mu(x)$ being four ``well behaved'' functions (of the $x^\mu$) that are seen as the parameters of the transformation $\mathscr{R}$ (similar as an
antisymmetric matrix $A$ is seen as the parameter of a Lorentz transformation).
If we introduce the notation:
\boxmath{
  R[r]^\mu_\nu(x) \eqdef \partial_\nu \{r^\mu\}(x)
 \quad \quad
  \tilde{R}[r]^\mu_\nu(x) \eqdef \partial_\nu \{(r^{-1})^\mu\}( r(x) )
} %end boxmath.
we have:
\boxmath{
  \tilde{R}^\mu_\alpha(x) R^\alpha_\nu(x) \eqdem \partial_\nu\{ (r^{-1}
  \circ r)^\mu (x) \}(x) \eqdem \partial_\nu\{ x^\mu \} \eqdem \delta^\mu_\nu
} %end boxmath.
\boxmath{
  R^\mu_\alpha(x) \tilde{R} ^\alpha_\nu(x) \eqdem \partial_\nu\{ (r
  \circ r^{-1}) ^\mu (x) \}(r(x)) \eqdem \delta^\mu_\nu
} %end boxmath.
With this, a general frame transformation for a $\phi(x)$, a $V^\mu(x)$,
a $T^{\mu \nu}(x)$, a $T_{\mu \nu}(x)$, etc, reads:
\boxmath{
  \mathscr{R}^s(r|\phi) ( \mathscr{R}^c(r|x) ) \eqdef \phi(x)
} %end boxmath.
\boxequa{general_frame_tsf_vec}{
  \mathscr{R}^{\onedot}(r|V) ^\mu ( \mathscr{R}^c(r|x) ) \eqdef R^\mu_\nu (x) V^\mu(x)
} %end boxequa.
\boxmath{
  \mathscr{R}^{\twodots}(r|T) ^{\mu\nu}  ( \mathscr{R}^c(r|x)) \eqdef R^\mu_\alpha(x) R^\nu_\beta(x) T^{\alpha \beta} (x)
} %end boxmath.
\boxmath{
  \mathscr{R}_{\twodots}(r|T)_{\mu\nu}  ( \mathscr{R}^c(r|x) ) \eqdef \tilde{R}_\mu^\alpha(x) \tilde{R}_\nu^\beta(x) T_{\alpha \beta} (x)
} %end boxmath.
For all the upper definitions, we have the important composition group property:
\boxmath{
  \mathscr{R}(r_1 | \mathscr{R}(r_2 | whatever )) \eqdem
  \mathscr{R}(r_1 \circ r_2|
  whatever )
} %end boxmath.
We can note that for these, a Lorentz transformation $\mathscr{L}$ is a
particular case of a $\mathscr{R}$ if we take:
\boxmath{
  r^\mu(x) \eqdef (\myexp{A \eta})^\mu_\nu x^\nu
} %end boxmath.

If, by using a $g_{\mu \nu}(x)$, we define the length of of a $V^\mu(x)$ with:
\boxmath{
  l[g](V)(x) \eqdef g_{\mu\nu} (x) V^\mu(x) V^\nu(x)
} %end boxmath.
with the upper definitions ($\ref{general_frame_tsf_vec}$), it is easy
to show that it is invariant by a general frame transformation:
\boxmath{
  l[ \mathscr{R}_{\twodots}(r|g) ]( \mathscr{R}^{\onedot}(r|V) )( \mathscr{R}^c(r|x) ) \eqdem l[g](V)(x)
} %end boxmath.

\subsection {Spinor transformation? Vierbein}
Concerning a spinor transformation $\mathscr{R}^\circ$, we have a slight problem since we have no
obvious extension of the Lorentz transformation $\mathscr{L}^\circ$  ($\ref{exp_D_A}$) for
them. To handle the Dirac equation within general relativity, people
do a (not so appealing) compound
construction by introducing ``vierbein'' fields
$e_\mu^\alpha(x)$  bearing two indices of different kind concerning transformations. The vierbein fields are 
considered to be more fundamental than the
metric $g_{\mu \nu}(x)$, and  in fact define the metric with:
\boxequa{R_metric}{
  g_{\mu \nu} (x) \eqdef e_\mu^\alpha(x) e_\nu^\beta(x) \eta_{\alpha \beta}
} %end boxequa.
 Are introduced also co-vierbein fields $\tilde{e}^\mu_\alpha(x)$
such that:
\boxmath{
  e_\mu^\alpha(x) \tilde{e}^\mu_\beta(x) \eqdef \delta^\alpha_\beta
\quad \quad
  \tilde{e}^\mu_\alpha(x) e_\nu^\alpha(x) \eqdef \delta^\mu_\nu
} %end boxmath.

\subsection {Local (or internal) Lorentz transformation $\mathscr{L}_I$}
Concerning transformations for $e_\mu^\alpha(x)$, the $\mathscr{R}$
one is applied for the $\mu$ down index, but an ``internal''
or ``local'' Lorentz one $\mathscr{L}_I$ is applied for the $\alpha$ upper one. The
$\mathscr{L}_I$ is defined as:
\boxmath{
  \mathscr{L}_I^c(A|x)^\mu \eqdef x^\mu
} %end boxmath.
(then no effect on $x^\mu$), and on a $\psi(x)$:
\boxmath{
  \mathscr{L}_I^{\circ}(A|\psi) ^\alpha (x) \eqdef (\myexp{\Gamma[A]})^\alpha_\beta \psi^\beta(x)
} %end boxmath.
We note then the important difference with a $\mathscr{L}$; the
$x^\mu$ are not changed with a $\mathscr{L}_I$ transformation.

\subsection {Dirac equation with vierbein}
The Dirac equation on ``curved spacetime'' is now written:
\boxequa{dirac_vierbein}{
   \tilde{e}^\mu_\alpha(x) i \gamma^\alpha \partial_\mu \psi(x) =
   \frac{mc}{\hbar} \psi(x)
} %end boxequa.
Someone can check that this equation is invariant by both the
$\mathscr{R}$ and $\mathscr{L}_I$ transformations with:
\boxmath{
  \mathscr{R}^s(r|\psi) ^\alpha ( \mathscr{R}^c(r|x) ) \eqdef \psi^\alpha(x)
} %end boxmath.
\boxmath{
  \mathscr{L}_I^{\circ}(A|\psi) ^\alpha (x) \eqdef (\myexp{\Gamma[A]})^\alpha_\beta \psi^\beta(x)
} %end boxmath.

\boxmath{
  \mathscr{R}_{\onedot}(r|e^\alpha)_\mu ( \mathscr{R}^c(r|x) ) \eqdef
  \tilde{R}_\mu^\nu(x) e_\nu^\alpha(x)
} %end boxmath.
\boxmath{
  \mathscr{L}_I^\circ(A|e_\mu) ^\alpha (x) \eqdef (\myexp{A\eta})^\alpha_\beta e_\mu^\beta(x)
} %end boxmath.
\boxmath{
  \mathscr{R}^s(r|e_\mu)^\alpha ( \mathscr{R}^c(r|x) ) \eqdef
  e_\mu^\alpha(x)
} %end boxmath.
\boxmath{
  \mathscr{L}_I^s(A|e^\alpha)_\mu (x) \eqdef e^\alpha_\mu(x)
} %end boxmath.

\boxmath{
  \mathscr{R}^{\onedot} (r|\tilde{e}_\alpha)^\mu ( \mathscr{R}^c(r|x) ) \eqdef
  R^\mu_\nu(x) \tilde{e}^\nu_\alpha(x)
} %end boxmath.
\boxmath{
  \mathscr{L}_{I \circ} (A|\tilde{e}^\mu) _\alpha (x) \eqdef
  (\myexp{-A \eta})_\alpha^\beta \tilde{e}_\beta ^\mu (x)
} %end boxmath.
\boxmath{
  \mathscr{R}^s(r|\tilde{e}^\mu)_\alpha ( \mathscr{R}^c(r|x) ) \eqdef
  e^\mu_\alpha(x)
} %end boxmath.
\boxmath{
  \mathscr{L}_{I s}  (A|\tilde{e} _\alpha)^\mu (x) \eqdef \tilde{e}_\alpha ^\mu (x)
} %end boxmath.

To play with the upper transformations over the vierbein and
co-vierbein fields, we can show that with the
metric defined by ($\ref{R_metric}$)  we have:
\boxmath{
  \mathscr{R}_{\twodots}(r|g)_{\mu \nu} (x) \eqdem
  \mathscr{R}_{\onedot}(r|e^\alpha)_\mu ( x)
  \mathscr{R}_{\onedot}(r|e^\beta)_\nu ( x)
  \eta_{\alpha \beta}
} %end boxmath.
and:
\boxmath{
  \mathscr{L}_{I s}(A|g)_{\mu \nu} (x) \eqdef g_{\mu \nu}(x) \eqdem
  \mathscr{L}_I^\circ(A|e_\mu)^\alpha (x)
  \mathscr{L}_I^\circ(A|e_\nu)^\beta (x)
  \eta_{\alpha \beta}
} %end boxmath.

The compound construction ($\ref{dirac_vierbein}$) is probably related to the fact that, put
all together, it makes no sense to attempt to mix the Dirac
equation, related to the ``quantum world'', with maths related to the
macroscopic world modeled with general relativity, and this without having
unified in first place the ideas related to these
two worlds. (A ``quantum gravity'' theory will probably be based on only
one single ``crystal clear'' transformation).

By using the Majorana representation, the equation
($\ref{dirac_vierbein}$) can be written:
\boxmath{
   h^{\mu \alpha} (x) \shorterquad \tilde{S}_\alpha \xi \shorterquad \partial_\mu \psi(x) =
   \frac{mc}{\hbar} \psi(x)
} %end boxmath.
\boxmath{
   h^{\mu \alpha}(x) \eqdef \tilde{e}^\mu_\beta(x) (\eta^{-1})^\beta_\alpha
} %end boxmath.
which is similar to ($\ref{h_tilde_Dirac_equation}$) but with the
$h^{\mu \alpha}(x)$ fields being now not constant and transforming in a
mixed way for each index (with $\mathscr{R}$ for the first one and
$\mathscr{L}_I$ for the second one). We can reach, here too, the
conclusion that this equation is not so natural from a group point of view and
something as ($\ref{h_Dirac_equation}$) or ($\ref{dirac_Sigmas_xi}$)
would be better.

\end{appendices}

\end{NoHyper}

%/////////////////////////////////////////////////////////////
%/// bib: /////////////////////////////////////////////////////
%/////////////////////////////////////////////////////////////


\begin{thebibliography}{10}

\bibitem{ref_iz}
  C.Itzykson, J.B.Zuber
  \textit{Quantum Field Theory} (McGraw-Hill International Book Company, 1980)

\end{thebibliography}
\end{document}